\documentclass[sigconf]{acmart}

\usepackage{tikz}
\usepackage{amsmath}

\usepackage{xspace}

\usepackage{filecontents}

\usepackage{multirow}

\usepackage{subcaption}
\usepackage{filecontents}
\usepackage{ifthen}
\usepackage[normalem]{ulem} 
\usepackage{amssymb}

\newboolean{showedits}
\setboolean{showedits}{true} 
\ifthenelse{\boolean{showedits}}
{
	\newcommand{\del}[1]{\textcolor{red}{\sout{#1}}} 
}{
	\newcommand{\del}[1]{} 
	
}

\newboolean{showcomments}
\setboolean{showcomments}{true}
\newcommand{\id}[1]{$-$Id: scgPaper.tex 32478 2010-04-29 09:11:32Z oscar $-$}

\ifthenelse{\boolean{showcomments}}
{\newcommand{\nbc}[3]{
 {\colorbox{#3}{\bfseries\sffamily\scriptsize\textcolor{white}{#1}}}
 {\textcolor{#3}{\sf\small$\blacktriangleright$\textit{#2}$\blacktriangleleft$}}}
 }
{\newcommand{\nbc}[3]{}
 \renewcommand{\del}[1]{} 
 }

\definecolor{ibcolor}{rgb}{0.4,0.6,0.2}
\definecolor{sfcolor}{rgb}{0.2,0.2,0.9}
\definecolor{gncolor}{rgb}{0.8,0.5,0.3}
\definecolor{pwcolor}{rgb}{0.6,0.0,0.6}
\definecolor{cycolor}{rgb}{0.0,0.8,0.2}

\usepackage{wasysym}

\definecolor{todocolor}{rgb}{0.9,0.1,0.1}

\author{Antoine Riard}
\email{antoine.riard@gmail.com}
\affiliation{\institution{}}

\author{Gleb Naumenko}
\email{naumenko.gs@gmail.com}
\affiliation{\institution{}}

\setcopyright{none}

\renewcommand\footnotetextcopyrightpermission[1]{} 

\begin{document}
\fancyhead{}
\title{Time-Dilation Attacks on the Lightning Network}


\graphicspath{{pictures/}}

\begin{abstract}

Lightning Network (LN) is a widely-used network of payment channels enabling faster and cheaper Bitcoin transactions. In this paper, we outline three ways an attacker can steal funds from honest LN users. The attacks require dilating the time for victims to become aware of new blocks by eclipsing (isolating) victims from the network and delaying block delivery. While our focus is on the LN, time-dilation attacks may be relevant to any second-layer protocol that relies on a timely reaction.

According to our measurements, it is currently possible to steal the total channel capacity by keeping a node eclipsed for as little as 2 hours. Since trust-minimized Bitcoin light clients currently connect to a very limited number of random nodes, running just 500 Sybil nodes allows an attacker to Eclipse 47\% of newly deployed light clients (and hence prime them for an attack). As for the victims running a full node, since they are often used by large hubs or service providers, an attacker may justify the higher Eclipse attack cost by stealing all their available liquidity.

In addition, time-dilation attacks neither require access to hashrate nor purchasing from a victim. Thus, this class of attacks is a more practical way of stealing funds via Eclipse attacks than previously anticipated double-spending.

We argue that simple detection techniques based on the slow block arrival alone are not effective, and implementing more sophisticated detection is not trivial. We suggest that a combination of anti-Eclipse/anti-Sybil measures are crucial for mitigating time-dilation attacks.

\end{abstract}

\maketitle

\section{Introduction}
\label{sec:intro}

Bitcoin is a peer-to-peer electronic cash system, which solves the double-spend problem with a trust-minimized architecture by letting everyone verify all transactions \cite{Nakamoto2008Bitcoin}. As of Nov. 2019, the system operates over at least 60,000 nodes\footnote{From https://luke.dashjr.org/programs/bitcoin/files/charts/software.html} simultaneously running Bitcoin protocol software, not including the many more users of custodial services and trusted solutions.

Public auditability of the Bitcoin transaction history is the foundation of removing trusted third parties. The drawback of public auditability is a constraint on the transaction throughput. \textit{Second-layer} protocols on top of Bitcoin were designed to overcome this limitation \cite{poon2016lightning, back2014sidechains, lerner2015rsk}. For example, in Lightning Network (LN) \cite{poon2016lightning}, solving the double-spend problem by shifting it to be a private matter (as opposed to being solved \textit{on-chain}).

Second-layer protocols introduce new assumptions when compared to the original Bitcoin threat model. In this work, we explore how LN users may be subjected to a risk of having their funds stolen, once their Bitcoin nodes are eclipsed from honest nodes.

At high level, we exploit the requirements to monitor the Bitcoin blockchain and to detect relevant transactions in a timely manner. Per \textit{time-dilation attacks}, a malicious actor slows down block delivery to the victim and then finalizes an expired state of the Lightning channel on-chain, before a victim can notice.

For a non-infrastructure attacker eclipsing full nodes is difficult, but definitely not impossible, as demonstrated by prior work \cite{Heilman2015Eclipse, Apostolaki2017Hijack, Tran2019Erebus}. Since full nodes in the LN are often used by hubs (or big service providers), we will show that an attacker may justify the high attack cost by stealing their aggregate liquidity during one short (several hours) Eclipse attack.

At the same time, we will demonstrate that Eclipse attacks are easier to carry out against the many LN users whose wallets rely on \textit{light client protocols} to obtain information from the Bitcoin network, and light client implementations are currently more vulnerable to Eclipse attacks then full nodes.

If an attacker has a payment channel to a victim, and a victim is eclipsed, the remaining attack is only a matter of time (hours to days), and the attack success rate is approximately 93\%. This makes our attacks \textit{as difficult as Eclipse attacks} in practice. When combined with the fact that time-dilation attacks require neither hashrate access nor purchasing from a victim, time-dilation becomes the most practical way of stealing funds via Eclipse attacks, compared to the well-known double-spending of eclipsed nodes.

The problem can't be addressed by simply detecting the slow block arrival due to the uneven intervals between mined blocks. More advanced detection-based measures should be deployed carefully, considering a number of trade-offs: the effect of false positives and a chosen recovery strategy on payment channels and Bitcoin in general. Mitigations to the time-dilation attacks should be built around strong anti-Eclipse measures.

The paper is structured as follows:
\begin{itemize}
\item We provide the background required to understand Bitcoin and the Lightning Network: their advantages, limitations and assumptions in Section \ref{sec:background}.
\item We discuss the preparation required for time-dilation attacks including eclipsing a victim's Bitcoin node and mapping a Bitcoin node to a Lightning node in Section \ref{sec:attacks_prep}.
\item We define our threat model, how to launch time-dilation, why it is so difficult to mitigate it by simply observing slow block arrival, and three ways of exploiting time-dilation in Section \ref{sec:exploiting_td}.
\item We discuss the optimal strategy for an attacker to exploit time-dilation considering the already implemented stale tip detection in Bitcoin Core, and measure the attack cost and the gain from the attacks in Section \ref{sec:evaluation}.
\item We suggest various measures to raise the bar for setting up time-dilation attacks on LN and  discuss the trade-offs of sophisticated detection-based measures in Section \ref{sec:countermeasures}.
\item We suggest a list of open questions for further discussion and research in Section \ref{sec:discussion}.
\item We discuss how our work complements the prior research on the security of Bitcoin and Lightning in Section \ref{sec:related_work}.
\end{itemize}

\section{Background}
\label{sec:background}

\subsection{Bitcoin Base Layer}
The primary goals of the Bitcoin system are relaying and validating financial transactions. Bitcoin solves the double-spending problem by organizing transactions into a sequence of blocks.

A transaction in Bitcoin is unconfirmed until it is included in a valid block. Then, the number of blocks created on top of that block represents the degree of confirmations the transaction has.
This confirmation indication works largely due to the Bitcoin built-in incentive system: mining a block is a difficult and expensive task, which may result in a reward.

The more confirmations a transaction has, the more confident the receiver is that the transaction is unlikely to be reverted. We use \textit{unlikely} because absolute transaction finality in Bitcoin does not exist by design.

However, the incentives are aligned in a way that reverting a larger number of blocks becomes more and more unprofitable under a fundamental Bitcoin assumption. This assumption is that a fraction of malicious mining power does not exceed 50\% in the long run (usually roughly defined as several hours to days).

Mining a Bitcoin block mainly consists of two phases: assembling a valid sequence of transactions and finding a random nonce, which would satisfy the Proof-of-Work \cite{back2002hashcash, Dwork2003PoW} algorithm requirements.

The Proof-of-Work difficulty adjustment rule makes sure the expected time of producing a block is 10 minutes on average.
These 10-minute average intervals, as well as an upper bound of the block size, are used for a number of reasons, related to security and scalability.

These rules have two negative consequences. First, they make Bitcoin transactions potentially expensive: competition for the block space creates a transaction fee market, which may drive fees up. Second, they make transactions slow: as explained above, in most cases confirming a transaction requires waiting for at least one block. Both of these problems become more apparent when more transactions are happening on the Bitcoin blockchain.

\subsection{Off-Chain Scaling}

To address these issues, off-chain scaling constructions were proposed \cite{Gudgeon2019L2}. These constructions are usually based on the techniques enabled by the Bitcoin scripting language, \textit{Script}. They are often referred to as \textit{Layer 2} because they operate on top of Bitcoin on-chain transactions (referred to as \textit{Layer 1}).

The security of off-chain protocols differs from the security of the Bitcoin protocol because at least one of the following holds:
\begin{itemize}
\item these protocols introduce an extra assumption on trusting third parties (e.g., a federation of operators \cite{back2014sidechains})
\item users are assumed to react in a timely manner to base layer updates \cite{poon2016lightning}
\end{itemize}

In Section \ref{sec:lightning_network}, we further discuss the second assumption (relevant for the LN), and later use it as a basis for the attacks we demonstrate.

Since Lightning Network heavily uses advanced transaction types, we will now explain the internals of Bitcoin transactions.

\subsection{Transactions in Bitcoin}

Transactions in Bitcoin consist of inputs, unlocking scripts, and outputs.
Inputs indicate which funds are being spent by a transaction. Unlocking scripts provide data required to verify that a spender indeed is authorized to access inputs.
Outputs define how the funds can be spent further, effectively defining who owns the funds and under which conditions.

In a simple scenario, a payee sends their public key to the payer, and the payer uses that as a transaction output. When a transaction is included in the blockchain, a payee can be sure they have access to those funds.

Every transaction may have multiple inputs and multiple outputs, and they don’t have to directly map one to each other.

Outputs can be spent via a simple digital signature or more complex conditions. For example, revealing a preimage for a pre-defined hash. These are called \textbf{Hash Time Locked Contracts (HTLCs)}. As the name suggests, a HTLC is built up from two primitives: a timelock and a hashlock. The contract semantics of an HTLC can be understood as “if a preimage $P$ is provided such $hash(P) == H\_lock$, before timelock expiration $T$, allow spending". These primitives are provided by the Bitcoin scripting language.

We will now discuss payment channel constructions and LN, based on these advanced spending conditions.

\subsection{Lightning Network}
\label{sec:lightning_network}

The high-level idea of payment channels was first suggested \cite{Nakamoto2013Channel} by the creator(s) of Bitcoin in 2011: cache transactions between the peers (payer and payee) instead of committing every transaction to the Bitcoin blockchain. Even though the described design was not secure, the high-level idea has since evolved and payment channels are now used for off-chain scaling.

The most widely used system based on Bitcoin payment channels is the Lightning Network \cite{poon2016lightning}: independent payment channels form a network, where users transact bidirectionally with other members of the network (via multi-hop payments).

Payment channels for the LN can be created after an out-of-band negotiation where two users decide that it makes sense for them to use channels instead of submitting every transaction on-chain. However, the software is often enabled to open channels without any negotiation.

Since the LN enables multi-hop payments, another common way to join the network is to use an on-demand service for the channel creation: create a channel to a hub, which would allow transacting with other users reachable (potentially, indirectly) via that hub.

LN uses a modified Poon-Dryja revocation mechanism \cite{poon2016lightning} to enable \textbf{bidirectional channels with unlimited lifetime}. At a high-level, proceeding with the new state reveals the revocation secret, which makes the previous state invalid.

Poon-Dryja payment channels are opened when a funding transaction, a 2-of-2 multi-signature contract between Alice and Bob, is submitted on-chain. This design enables a channel close with any outcome if both of the parties are online and cooperating.

To ensure the security of the funds even if the counterparty is irresponsive, transactions spending the multisig funding output ("commitment") must always be valid and ready to broadcast. Therefore at every state update, encoded by a new commitment transaction, signatures must be exchanged. If Alice initiates the update, she sends signatures for Bob's new transactions. Bob then revokes his previous set of transactions and sends his signatures to Alice for her new transactions.

This transaction asymmetry and the structure of non-commitment transactions (which we show in detail below) allow every party to unilaterally close a channel without further interactivity. At the same time, they enable punishment by the counterparty, if channel closing is dishonest.

Committing to an outdated state on-chain by a malicious actor is disincentivized by a \textit{punishment} time-window. During this time, an honest user can confiscate all funds of a malicious counterparty through a justice transaction. The time-window is enforced directly via relative timelocks \cite{Friedenbach2015Csv}.

Every state in LN is enforced by 6 types of transactions \cite{Bolt2019Format}, as seen from Alice's viewpoint.
Alice has 3 transactions, fully-countersigned by Bob and ready to broadcast:
\begin{itemize}
\item Commitment transaction, used by Alice to finalize the state. It has 4 types of outputs: Alice's balance, Bob's balance, offered HTLC, received HTLC. An offered HTLC allows locking a conditional payment flowing from Alice to Bob. A received HTLC allows a conditional payment flowing in the reverse, from Bob to Alice. The distinction enables bidirectional payment.
\item HTLC-Timeout, used by Alice to spend an offered output on her commitment transaction. It allows her to refund after timelock expiration.
\item HTLC-Success, used by Alice to spend a received output on her commitment transaction. It allows her to get paid by presenting a preimage before timelock expiration.
\end{itemize}

For every offered or received HTLC output, Alice must have a corresponding HTLC-Success or HTLC-Timeout transaction.

Bob may generate 3 single-signed (no need to update Alice's signatures) transactions in reaction to Alice behavior:
\begin{itemize}
\item Preimage transaction, used by Bob to spend an offered output on Alice's commitment transaction. Allows him to get paid by presenting a preimage before timelock expiration.
\item Timeout transaction, used by Bob to spend a received output on Alice's commitment transaction to refund himself after timelock expiration.
\item Justice transaction, used by Bob to spend any output belonging to Alice's revoked transaction. Allows him to confiscate Alice's funds by using previously revealed per-update revocation secret.
\end{itemize}

Since the channel is symmetrical, Bob holds his own commitment, HTLC-Timeout/HTLC-Success, on which Alice can generate her reaction transactions.

\textbf{Multi-hop payments} are enabled in LN by routing HTLCs across a path of nodes.

\begin{figure}[t]
\centering
\includegraphics[width=1.0\linewidth]{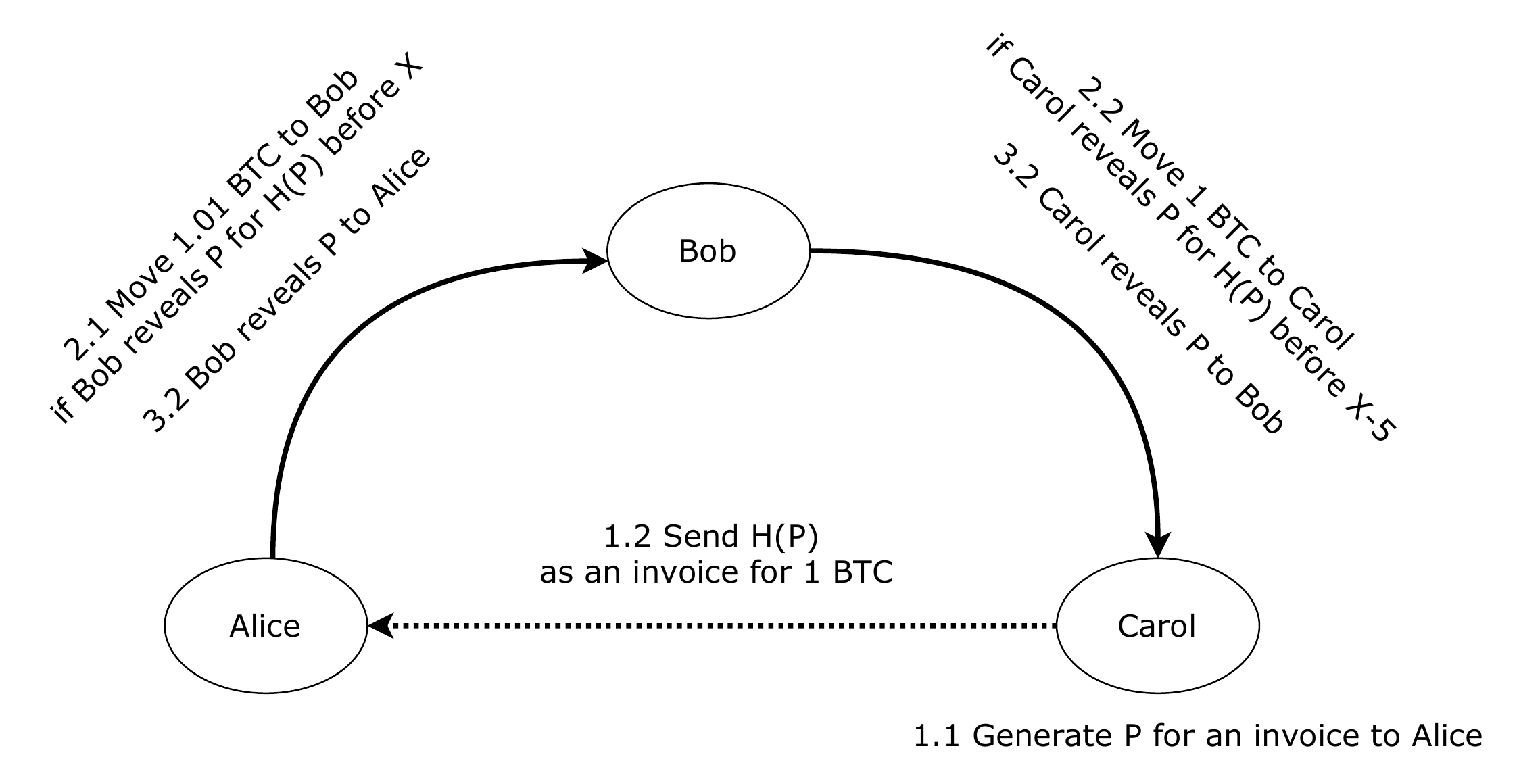}
\caption{Routing payments in Lightning Network. }
\label{fig:routing}
\end{figure}

While the payment is routed through the LN, the whole payment path shares a hashlock. A timelock is decreased at every hop from a payer to a payee.
Every multi-hop payment consists of three phases (see Fig. \ref{fig:routing}):
\begin{enumerate}
\item A payee sends to the payer an invoice containing a hash to a preimage chosen by the payee.
\item Route setup, when every party agrees with the next hop to add an HTLC on their local channel sequentially in every channel, ordered from the payer to the payee.
\item Settlement phase, where every party agrees with the previous hop to remove the HTLC, once the preimage is known to the channel participant from the payee side. The preimage is propagated all the way to the first channel in the chain.
\end{enumerate}

This sequence of decreasing timelocks enables a secure in-order HTLC settlement. An intermediate hop is always able to claim an incoming HTLC with a timelock less or equal to the outgoing one, or to cancel it after a state update. In other words, it makes an intermediate router protected from the loss of an HTLC.

The per-hop delay (timelock difference) used for this protection is called $cltv\_delta$. It is enforced by each intermediate hop during the route setup phase, at the reception of an incoming HTLC.

\subsection{Extra Assumptions}

The cost of scaling solutions based on payment channels is new assumptions, of which the following is relevant for our work: \textit{a user should always have access to the recent blockchain history and should be able to broadcast transactions, in the case of counterparty misbehavior}.

Fundamentally speaking, LN introduces new security parameters. Instead of measuring the finality of the transactions with confirmations only (number of blocks after inclusion in the blockchain), the security of payments in the LN should also be measured by the chosen timelocks. The longer funds are locked in a channel, the better chance an honest user has to act on the misbehavior of a counterparty and get their funds back from the channel. At the same time, it makes the protocol less flexible for an honest unilateral close, triggered by an irresponsive counterparty.

The required blockchain monitoring can be done via running a full Bitcoin node, by relying on a trusted third party or by using a \textit{light client}. We do not cover issues with trusting third parties in our work, because we focus on the non-custodial use.

In theory, partial third party trust can significantly increase the security, and, to the best of our knowledge, this method is used by several popular Lightning wallets. But again, it changes the threat model by introducing trust.

For example, if a Lightning wallet is based on secure open-source software but doesn't have strong Eclipse protection, the trusted node still can choose to steal funds via time-dilation. While it might not make sense against a single client, an exit scam from a wallet developer (usually operating the trusted node) stealing funds from all the channels is plausible, even if the software is secure otherwise. Thus, by focusing on a trust-minimized scenario we cover the security of these clients as well.

\subsection{Light Client Protocols}
\label{sec:light_clients}

Several protocols have been proposed to reduce the requirement of running a full node and still use Bitcoin with a fairly trust-minimized model. All of them use client-server architecture with multiple servers, assuming that at least one of the servers a client connects to is honest.

Light clients are often used as a Bitcoin blockchain processing backend on resource-constrained devices (like mobile phones). Understanding the security of these clients is important to evaluate the security of LN client implementations.

The first popular non-standardized implementation of a light client is \textbf{Electrum}. Per this protocol, Electrum light clients connect to Electrum servers. Electrum server must have access to the chain processing backend, usually co-located on the same machine with the server.

Electrum itself provides configurations with different trade-offs. For example, an Electrum user can connect their light clients to Electrum Personal Server software run by themselves, or connect to multiple reachable ElectrumX Servers run by someone else.

Electrum is currently used as a Bitcoin chain processing backend by one of the most popular Lightning wallets, \textit{Eclair}.

\textbf{BIP 157} is the most popular standardized light client protocol. Clients based on this protocol would connect to full nodes in the Bitcoin network, receive a compact representation of Bitcoin blocks (filters, as defined in another related standard, BIP 158), and, if a filter detects relevant transactions on the client-side, request a full block of transactions.

Neutrino is currently one of the most popular light client implementations, and it is based on BIP 157. Neutrino is used by at least \textit{Breez} and \textit{Wallet by Lightning Labs}).

We will now provide the background on the relevant attacks on the Bitcoin network and light clients, required to understand time-dilation attacks on the Lightning Network. We will cover the robustness of full Bitcoin nodes, Neutrino clients and Electrum clients, because these are the most widely used Bitcoin backends in the Lightning Network. We will not cover the security of other light client protocols (e.g., BIP 37) because their implementations are much less used and maintained.

\section{Attack Preparation. Eclipse and Node Mapping}
\label{sec:attacks_prep}

All Bitcoin nodes constitute a peer-to-peer network. Full Bitcoin nodes can be roughly split into two categories:
\begin{itemize}
\item reachable from most of the Internet and accepting inbound connections from other nodes
\item non-reachable nodes behind NATs and firewalls
\end{itemize}
Reachable nodes act as a backbone, allowing other nodes to join and relay transactions, blocks, and other necessary information.

As of March 2020, every Bitcoin Core node by default maintains up to 8 outbound connections to relay transactions, blocks, and network addresses of other nodes; and 2 extra connections to relay exclusively blocks. All connections in the Bitcoin network are bidirectional.
Connections relaying only blocks leak less information and are supposed to secure block relay.

Although outbound peer rotation has been discussed multiple times \cite{Naumenko2018Rotation, Pustogarov2014Rotation}, Bitcoin Core never deviated from the status quo approach. Thus, the topology is currently fairly static, and new outbound connections for an existing node are only made due to issues with existing connections.

Since the network is permissionless, it is naturally susceptible to certain attacks, which enable time-dilation. We consider two practical scenarios for time-dilation:

\textbf{C1.} Victim’s Bitcoin node is first eclipsed (isolated) as a part of a broader attack on the Bitcoin network, and an attacker attempts to find a corresponding Lightning node.

\textbf{C2.} Specific victim’s Lightning node (identified by IP and the channels) is targeted, and then an attacker attempts to locate and eclipse a corresponding Bitcoin node.

In both cases, an attacker would have to eclipse a victim's Bitcoin node and map a Lightning node to a Bitcoin node (often involving transaction origin inference), in different orders.

These attacks are relevant against LN users running their own full nodes or light clients for Bitcoin blockchain processing, instead of relying on third parties. We will now describe the relevant attacks targeting full nodes, and then discuss applying them to light clients in more detail.

\subsection{Eclipse Attacks on Full Nodes}

\begin{figure}[t]
\centering
\includegraphics[width=1.0\linewidth]{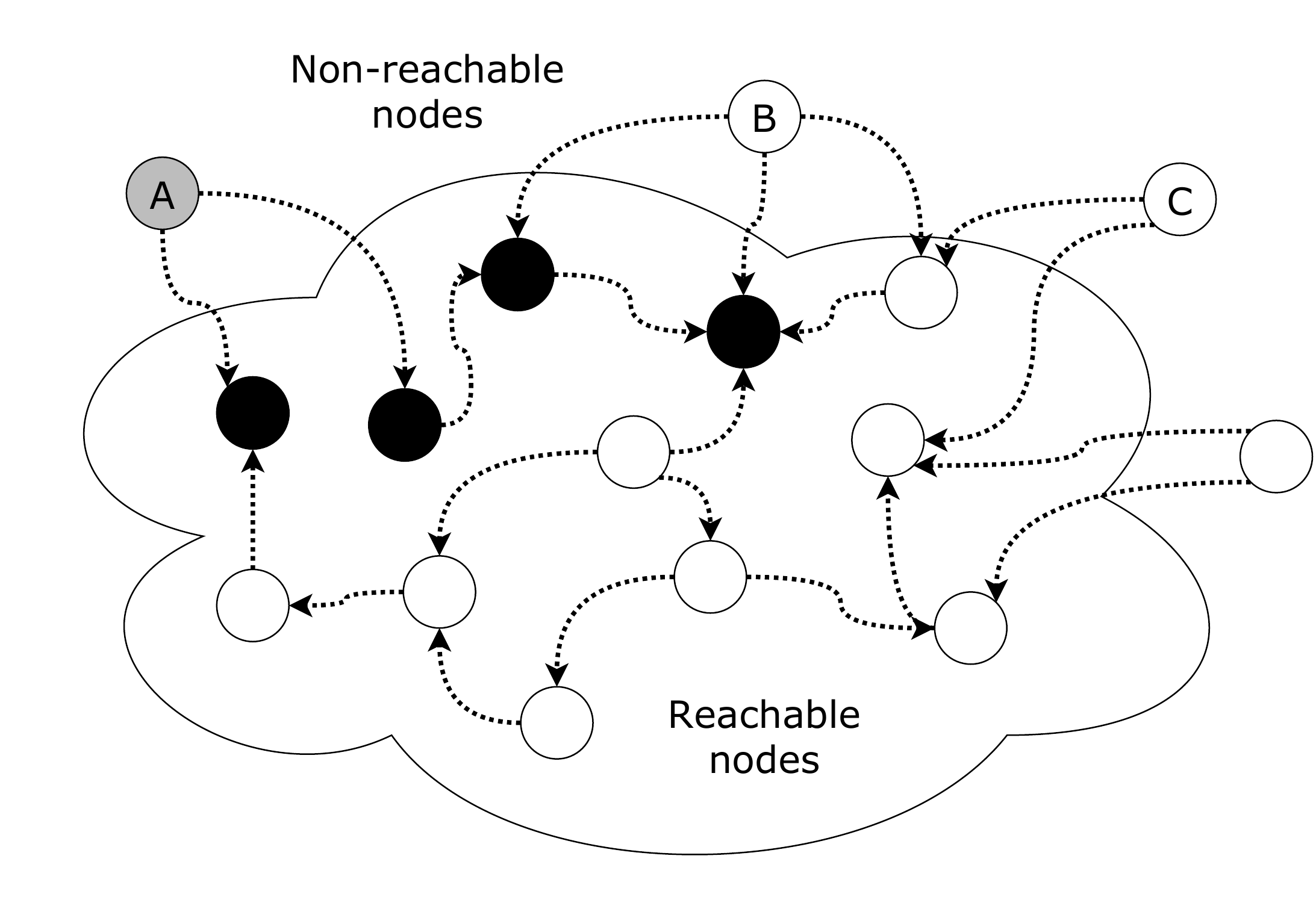}
\caption{Eclipse attack on the Bitcoin network. Only node $A$ is eclipsed, because all its connections lead to the attacker.}
\label{fig:eclipse}
\end{figure}

By definition, an Eclipse attack implies preventing a victim's node from communicating with other honest participants of the network. It is usually done by occupying all of the victim’s node connections by malicious nodes or pseudo-nodes. An attacker gains complete control over \textit{what} and \textit{when} a victim sends and receives from the network. This is a crucial requirement for performing time-dilation attacks. Fig. \ref{fig:eclipse} demonstrates an Eclipse attack from the topology perspective.

The first Eclipse attack on the Bitcoin network was demonstrated by Heilman et. al. \cite{Heilman2015Eclipse}. It is purely based on the high-level protocols of the Bitcoin network, namely, address management and relay logic.
Further research demonstrated that using a combination of the exploitation of the BGP protocol and Bitcoin’s address management can significantly reduce the eclipsing cost. Apostolaki et. al. \cite{Apostolaki2017Hijack} demonstrated that any network-level attacker (e.g., Autonomous System) can isolate a significant portion of nodes by hijacking less than 100 prefixes, although the attack can be detected.
Tran et. al. \cite{Tran2019Erebus} further shown that any Tier-2 Autonomous System can Eclipse most of the reachable Bitcoin nodes in an undetectable way.

The studied consequences of eclipsing a Bitcoin node include monetary (double-spending attacks, attacks on mining) and non-monetary (peer-to-peer layer deanonymization). Although Heilman et. al. briefly mentioned monetary consequences on second layer protocols \cite{Marcus2018Eclipse}, our work is the first to discuss these attacks in detail.

\subsection{Eclipse Attacks on Neutrino}
\label{sec:eclipse_neutrino}

It is crucial for BIP 157 light clients to be connected to Bitcoin full nodes providing the required server-side support.

The number of available honest serving nodes defines the security of light clients against Sybil attacks. The more honest nodes serve light clients, the more Sybil nodes an attacker needs to deploy to Eclipse a victim. The security of light clients can in theory be increased by connecting to regular full nodes and checking the tip with them, but this behavior is currently not implemented.

Thus, for a given number of attacker Sybil serving nodes ($N_a$), and a number of honest serving nodes ($N_h$), and the number of outgoing connections every honest node maintains ($C$), the probability of a successful Eclipse attack can be then measured as:

\begin{equation}
P_{E} = (\frac{N_a}{N_h+N_a})^{C}
\end{equation}

To estimate the cost of eclipsing honest nodes via a trivial Sybil attack today, we collected a list of available serving nodes from the Bitcoin DNS seed servers. This is the same procedure any new node in the network follows to learn about nodes in the network during the initial bootstrap. After getting a subset of this list, light clients usually choose 8 peers at random.

Currently, only one of the Bitcoin implementations (\textit{btcd}) has released a build with the server-side support for these light clients. There are only around 30 reachable nodes in the network that run this implementation. We also found 20 nodes running a custom version on Bitcoin Core, based on the work-on-progress implementation of server-side BIP 157 support.

According to the formula above, spawning just 500 Sybil nodes with server-side BIP 157 support would trivially eclipse a random 47\% of newly deployed or restarted light clients.

To increase the probability of success for the attack, an attacker would have to either create more Sybil server nodes or reduce available honest server nodes. The former requirement makes it expensive to Sybil attack the whole Bitcoin network, but quite practical to attack the subset of nodes using BIP 157/158, because this feature is supported by so few honest nodes. The latter can be achieved via DoS.

Additionally, an attacker might exploit that neither Neutrino nor btcd implement countermeasures discussed in \cite{Heilman2015Eclipse, Tran2019Erebus}, and those used in Bitcoin Core. For example, they don’t employ the following methods:
\begin{itemize}
\item peer selection diversification based on an Autonomous System a peer belongs too, to make it more difficult for an attacker to get victims exclusively connect to the attacker’s Sybil nodes
\item eviction on inbound connections when all the slots are occupied, to allow new nodes to connect to honest reachable nodes even when their inbound capacity is exhausted.
\end{itemize}

Lack of these and other countermeasures in light clients allow more sophisticated Eclipse attacks to succeed even at a lower cost.

\subsection{Verifying Eclipse}

Once an attacker suspects that a victim Bitcoin node is eclipsed, they should verify that a node does not have another form of access to the Bitcoin network.

The easiest way to verify a node is eclipsed is \textbf{transaction probing} based on transaction relay protocols. An attacker chooses a random transaction they received from the network, and don't relay it to the eclipsed node through any of the connections under control. Then, if a victim’s node announces that transaction to an attacker, it means there is still a link between a victim and an honest part of the network.

It would be more difficult if a node is connected to an external source of \textit{blocks}, which does not relay transactions. The proposed methodology would identify eclipsing only the transaction relay aspect of the peer-to-peer communication, while these attacks require eclipsing all links relaying blocks.

In this case, an attacker would have to apply \textbf{block probing}: delaying a block delivery through all links to the victim, and observing whether a victim relays that block to the attacker nodes. The only problem with this approach is that blocks arrive much less often than transactions. Thus, if the probing demonstrated that the victim is still not eclipsed, the next attempt will be possible no earlier than in 10 minutes on average (as opposed to every second with transactions).

Sometimes a victim may have an irregular source of blocks or transactions (e.g., Blockstream Satellite service). In this case, \textbf{a combination of block and transaction probing} would help an attacker to timely identify that this is the case, deduce what kind of external service a victim is using and whether they are capable of disrupting this service. This would ultimately help the attacker to choose a better strategy for proceeding with the attack.

\subsection{Mapping Nodes}
\label{sec:mapping_nodes}

To launch a time-dilation attack, a malicious actor also has to map a victim's Bitcoin node to a Lightning node.

The easiest mapping technique is correlating Bitcoin and Lightning nodes that \textbf{operate under the same IP}. To measure how many users run their nodes under the same IP, we scraped IP addresses of the Bitcoin nodes over a week. Then we correlated them with the list of lightning nodes with advertised channels.

We were able to gather a list of 4,500 Lightning nodes and 52,000 Bitcoin nodes and found 982 matches by IP.
Almost half of the Lightning nodes were represented by an onion address, making them even less likely to be traceable by this methodology. Only two pairs of nodes shared the onion address.
These numbers do not include Lightning nodes with \textit{private} (not advertised) channels and non-listening Bitcoin nodes.

If Bitcoin and Lightning nodes \textbf{operate under different IPs}, an attacker would have to apply the following heuristics.

In case \textbf{C1}, an attacker would have to find which Lightning channel funding transactions originated from the victim’s eclipsed Bitcoin node, and map those transactions to channel announcements in the Lightning network.
The most straightforward approach is to apply transaction origin inference against Lightning-related transactions coming from the victim’s Bitcoin node. It can provide precise results because an attacker can analyze all the relevant messages coming from/to the victim node, acting as a Man-in-the-Middle between a victim and the honest part of the network.

Alternatively, an attacker can withhold a block from the victim’s eclipsed Bitcoin node, and look for the nodes in the Lightning Network which don’t accept and relay some of the channel announcements, which become valid \textit{within the withheld block}.

For case \textbf{C2}, an attacker would have to:
\begin{enumerate}
\item Deploy Sybil nodes in the Bitcoin network, both connecting to honest nodes and accepting connections from them
\item Apply transaction origin inference to the relay of the Bitcoin transactions corresponding to the victim’s channels
\end{enumerate}

In both cases, the approach involving transaction origin inference might take days or even weeks.

For \textbf{C1}, the upper bound on time is set by the channel's lifetime, which can be unlimited in the LN. If a victim never commits any channels on-chain, it is impossible to map their nodes. In practice, channels do get closed, although the lifetime may vary from hours to weeks, and an average channel age is currently 319 days \footnote{As shown by https://1ml.com/statistics on 2020-05-15}.

Alternatively, an attacker can proactively open low-value channels with a victim and infer from them, if the victim's LN implementation is configured accordingly (often enabled by default). This technique would require an attacker to spend the minimum cost of opening a channel per probe.

For \textbf{C2}, the upper bound is set by the time it takes to passively accept enough incoming connections from honest nodes. It may take days because honest nodes rarely add new peers (only when an existing peer is disconnected), and forcing reconnections requires attack capabilities beyond our threat model. Proactive connection to victims is not enough, because in Bitcoin Core transactions are relayed to inbound connections slower to make Sybil-based spying less effective.

We will now discuss the known techniques and the feasibility of the transaction origin inference required for mapping nodes.

\subsection{Transaction Origin Inference on Full Nodes}

Transaction origin inference means finding a Bitcoin node, from which a particular transaction was initially relayed. This would allow linking a transaction to a particular IP address, assuming a transaction sender uses their own node to submit transactions. We anticipate that this is a fair assumption for LN users who prefer a trust-minimized model compared to relying on a third-party.

It is possible that a transaction was relayed via a proxy node or Tor, in which case it would trigger a false positive observation, but this is currently not the default behavior and not the general case \footnote{Only 5 of 17 popular wallets listed at bitcoin.org have the Tor feature as of 2020-01-10}.

Transaction origin inference was previously explored in \cite{Biryukov2014Deanonymisation, Neudecker2016Timing, Grundmann2019Exploiting, Miller2015DiscoveringB, Delgado2019TxProbe}. Most of the demonstrated attacks are a form of a Sybil attack and use the first-spy estimator.
The first-spy estimator technique relies on the assumption that the node which announces a transaction earlier than other nodes, is likely to be an originating node to the transaction or is directly connected to the originating node \cite{Fanti2017Dandelion}.

Related work demonstrated that it is currently possible to infer transaction origin at high accuracy \cite{Naumenko2019Erlay, Fanti2017Dandelion} with a moderate network of Sybils.

\subsection{Transaction Origin Inference on Neutrino}

As we explained before, network-level transaction deanonymization in the suggested threat model usually relies on first-spy estimation: establishing multiple connections to honest nodes in the network and analyzing the messages coming from those nodes.

Bitcoin Core employs several techniques to obfuscate transaction flow across the network. These include:
\begin{enumerate}
\item random “diffusion” delays before announcing a transaction
\item increased diffusion delay for inbound connections
\item shared diffusion delay timer for all inbound connections
\item diverse node connectivity based on the IP ranges or Autonomous Systems
\end{enumerate}

None of these would work for Neutrino, because those light clients broadcast only their own transactions. Thus, it is enough for an attacker to make sure it has \textbf{at least one direct connection} from the Neutrino node of a victim, to infer the origin. In this case an attacker has to be sure that the victim runs a light client (and not a full node), which is currently trivial to infer from the peer-to-peer behavior of the victim.

Neutrino clients currently connect to a very limited number of public nodes (see Section \ref{sec:eclipse_neutrino}). Every Neutrino client chooses 8 nodes at random from the available pool of ~50 honest nodes serving filters per BIP 158.
An attacker with only 100 Sybil nodes can be sure that a victim is directly connected to a Sybil node at least once with a 97\% chance. This would allow the attacker to identify a source of a given transaction with very high accuracy and low cost.

\subsection{Attacks on Electrum Light Clients}

Robustness to Eclipse attacks and transaction origin inference of Electrum light clients depends on the chosen mode of operation.

If an Electrum user runs their own Electrum Personal Server or ElectrumX Server connected to their own Bitcoin full node, the user inherits the security from Bitcoin Core, partially described previously in this section.

If an Electrum user connects to ElectrumX Servers run by someone else, they face the same issues as Neutrino. A very low number of deployed servers\footnote{61 server, as listed at https://1209k.com/bitcoin-eye/ele.php as of 2020-04-13} and lack of strong anti-Sybil measures (compared to Bitcoin Core) make it easy to eclipse honest users.

\section{Time-dilation and the attacks}
\label{sec:exploiting_td}

In this Section we demonstrate the conservative threat model we chose, discuss the nature of time-dilation, and suggest three practical ways to steal money from LN channels.

\subsection{Threat Model}
\label{sec:threat_model}

First of all, we assume that an attacker can open a payment channel to a victim. Although it can be done both before and after eclipsing the victim, in our work we assume the former for simplification. The process of opening a channel is discussed in Section \ref{sec:lightning_network}.

We also make the following assumptions:
\begin{itemize}
\item Users run unmodified Bitcoin and Lightning node software.
\item The blockchain provides transaction safety based on confirmations: mining hashrate is stable and blocks are mined reliably.
\item The network of honest users forms a connected graph, except for a victim eclipsed by an attacker.
\end{itemize}

For simplicity we also assume that blocks are reliably relayed across nodes within seconds. We refer to the latest known block as "blockchain tip".

When it comes to the capabilities of an attacker, we consider:
\begin{itemize}
\item An attacker does not control any hash-rate.
\item An attacker can deploy hundreds of Sybil nodes with modified Bitcoin node software.
\item An attacker does not exceed the network-level capabilities discussed in the prior art on Eclipse attacks \cite{Heilman2015Eclipse, Tran2019Erebus, Apostolaki2017Hijack}.
\end{itemize}

This threat model allows an attacker to execute the underlying attacks (eclipsing, node mapping). An attacker then becomes capable of time-dilating a victim and stealing funds from their payment channels.

\subsection{Time-dilation}
\label{sec:td}

After a node is eclipsed (and there is a payment channel to a victim), an attacker has to perform time-dilation: slowing down block delivery to the victim’s Bitcoin node. Time-dilation is possible (can't be trivially detected) because, as we discussed in Section \ref{sec:background}, block mining is a Poisson process. For example, even though it is expected to see blocks every 10 minutes on average, seven blocks a day generally take longer than 30 minutes to be produced.

To time-dilate a victim, an attacker has to simply introduce a delay between receiving a block and feeding it to the victim. Since the victim is eclipsed and doesn't have an honest source of blocks, the attacker can decide when the victim receives a new block.

As of today, no dedicated countermeasure letting a victim distinguish between deferred block propagation from a random event is implemented in any of the Bitcoin client software. Furthermore, \textbf{countermeasures based on the delivery time alone can't be effective against time-dilation}: they either have high false positive rate or high false negative rate.

In other words, if these detections are triggered too often, and a node is configured to force close channels on this trigger, channels become less attractive economically, because they become very short-lived. There are also privacy issues: if emergency block fetching is triggered too easily (even by naturally slow blocks), it leaks the fetcher's privacy, which may enable more severe attacks.

If they are triggered not often enough, they allow an attacker to adapt to them (for example, by time-dilating at a pace which doesn't trigger them), so that the attack still can be launched undetectably, although it may take a little longer.

Thus, a good detection-based countermeasure implemented in the Bitcoin client should have a low true negative rate, but at the same time trigger a less radical (e.g., warning to a node operator) action. But even then, a victim would attempt to connect to the honest network again, which boils down to anti-Eclipse and anti-Sybil measures which could have been taken in the first place even without this trigger.

In Section \ref{sec:evaluation}, we demonstrate how stale tip detection in Bitcoin Core bounds an attacker in terms of the maximum time-dilation per block, and suggest an optimal attack strategy which makes stale tip detection ineffective against time-dilation.

Once the victim’s Bitcoin node is confirmed to be eclipsed and an attacker is able to slow down block delivery to that node, time-dilation attacks can be launched.

In the following descriptions, pseudonyms “Alice” and “Bob” represent users of the Lightning Network, "Mallory" and "Mallet" represent an attacker's entities.

\begin{table}[]
\begin{tabular}{l|l|l|l|l}
 & Implementation & CSV delta & CLTV delta & Timeout Policy\\
 \hline
 & C-lightning    & 144      & 14  & 7 \\
 & LND            & 144-2016 & 40  & 10 \\
 & Eclair         & 720      & 144 & 11 \\
 & Rust-lightning & 144      & 72  & 6
\end{tabular}
\caption{\label{tab:config_impl} Default timelocks (in blocks).}
\end{table}

We start by examining the scenario targeting the channel state finalization delay, the hardest to exploit in practice but the most studied so far. Then we explore more creative attacks targetting per-hop packet delay and packet finalization delays, the latter being much more practical than the other two.

\subsection{A1. Targeting Channel State Finalization}

This attack is structured similarly to the regular on-chain Bitcoin double-spend.

Let's say Alice and Mallory have a payment channel. The channel is configured with a CheckSequenceVerify\footnote{A delay (in blocks) timelocking the spending transaction based on the confirmation height of the spent transaction} \cite{Friedenbach2015Csv} timelock of $C$ blocks for contestation (see channel design in Section \ref{sec:lightning_network}). The default choice of $C$ in major LN implementations is summarized in \ref{tab:config_impl}.

To start exploiting an attack, Mallory should make Alice be $C$ blocks behind the actual tip of the blockchain by performing time-dilation. As a result, Alice's block height is pinned at $H-C$, where $H$ is the height of the actual latest block in the network.

Once the difference in heights is achieved, Mallory can double-spend Alice. To do so, Mallory negotiates with Alice a new state. Per this new state, Mallory pays Alice and receives something (in an irreversible way, like a physical or digital good) from Alice. Then Mallory commits the previous state on-chain with a preferred outdated state. The malicious revoked commitment transaction is settled on-chain at $H+1$.

Since the latest block Alice sees corresponds to time $H-C$, she won’t detect the channel revocation until reaching $H+1$. At that time, the honest network and Mallory are already at height $H+C+1$. The contestation period is expired for the rest of the honest network, and the malicious spend is fully valid.

\subsection{A2: Targeting Per-Hop Packet Delay}

This attack is based on exploiting the HTLC-based routing.

The attack starts with two lightning channels being opened: Mallory-Bob and Bob-Mallet. Bob enforces a $cltv\_delta$ (see Section \ref{sec:lightning_network}) of $M$ blocks on incoming HTLCs. We summarize how different LN implementations choose the $cltv\_delta$ in Table \ref{tab:config_impl}.
Mallory and Mallet eclipse Bob’s Bitcoin node and perform time-dilation until they gain a lead of $M+1$ blocks on Bob.

Once Mallory and Mallet have managed to be $M+1$ blocks ahead of Bob, they route a payment through him with a final timelock delta of $N$. On Bob-Mallet channel, HTLC timelock expires at $H+N$. On Mallory-Bob channnel, HTLC timelock expires at $H+M+N$, therefore satisfying Bob's $cltv\_delta$ of $M$. As before, $H$ is the height of the actual latest block in the network.

Once the actual Bitcoin blockchain tip is at height $H+M+N$, Mallet provides a required preimage to Bob  and gets from Bob a signature for a new state.

\begin{figure}[t]
\centering
\includegraphics[width=1.0\linewidth]{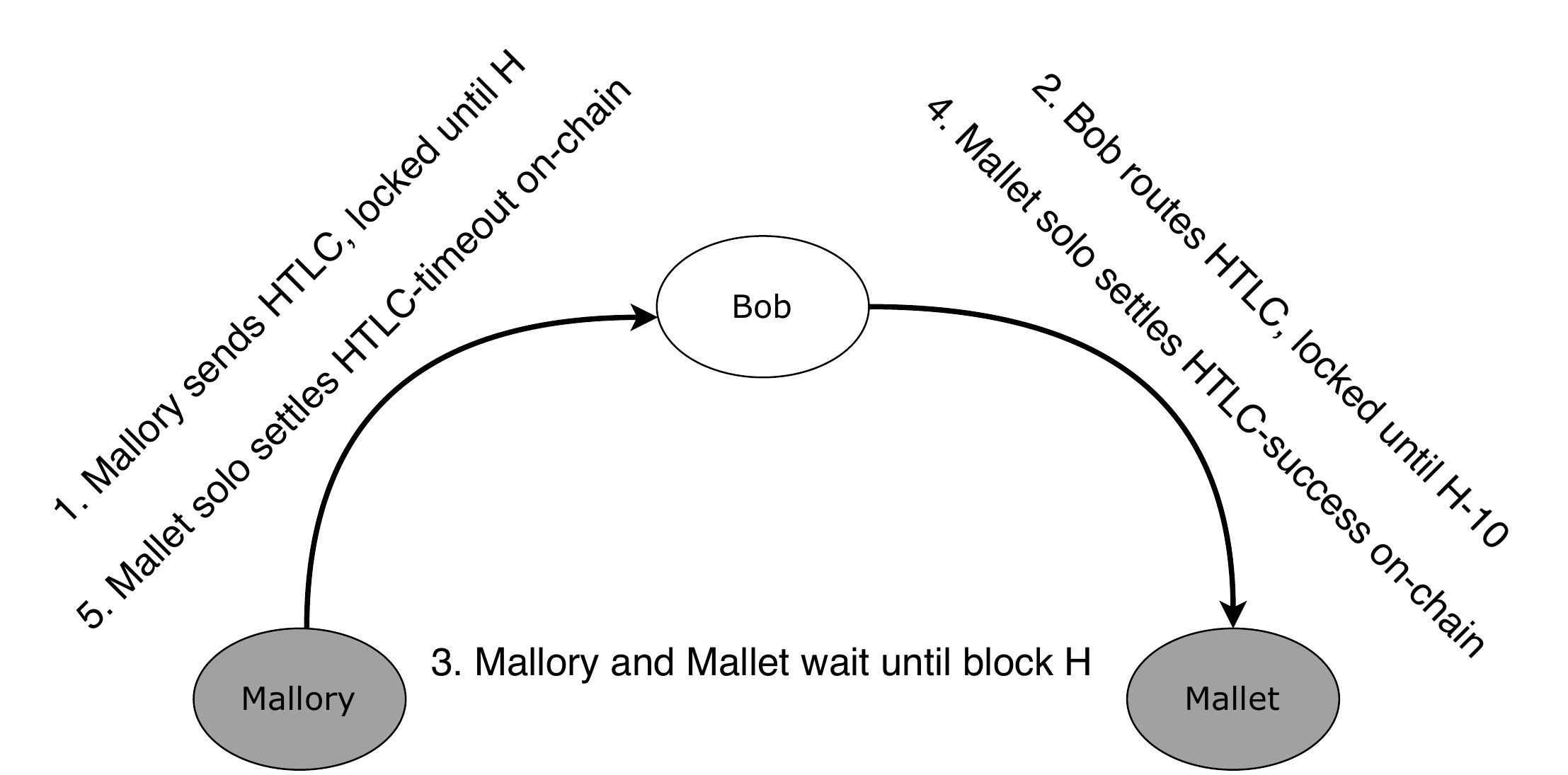}
\caption{Attack A2. Mallory and Mallet are attackers, Bob is a time-dilated victim. Steps 4-5 happen simultaneously.}
\label{fig:a2}
\end{figure}

At the same time, Mallory finalizes the state of her channel on the Bitcoin blockchain and broadcasts her HTLC-timeout transaction to get back the offered payment. This prevents Bob from re-negotiating the state of that channel via the preimage he just got against a full-payment to Mallet, making him effectively robbed. We summarize the attack in Fig. \ref{fig:a2}.

\subsection{A3: Targeting Packet Finalization}

This attack is based on exploiting the incoming HTLC safety delay on a channel. When a party knows the preimage for an incoming HTLC but the remote peer doesn't respond in a timely manner to update channel state, the party
will go on-chain to claim the incoming HTLC a few blocks before expiration.

Mallory, the attacker, starts by time-dilating Bob, the victim, by $I+1$ blocks (as specified in Table \ref{tab:timings_a3}). The attacker has to wait an extra block to avoid a broadcast race condition between an honest preimage transaction and a malicious HTLC-timeout.

At $H$, an HTLC is routed from Alice via Mallory to Bob and will expire at $H+N$, with $H$ actual latest block in the network, and $N$ final timelock delta. There is no collusion between Alice, an honest payer, and Mallory. Bob reveals the preimage to Mallory, and Mallory deliberately doesn't reply back to update channel state. When the blockchain tip reaches $H+N$ on the non-eclipsed network, Mallory broadcasts her commitment and HTLC-timeout transactions, therefore making revealed preimage invalid to claim offered HTLC on the Mallory-Bob channel.

Finally, when Bob reaches $H+N-I$ on his blockchain view, he attempts to claim the incoming HTLC by broadcasting his Preimage transaction. This one is going to be rejected by other network peers, the HTLC output has already been finalized by Mallory's HTLC-timeout transaction. Then, Mallory claims offered HTLC on Mallory-Alice channel, presenting Bob's preimage, therefore earning a routed payment for which she hasn't sent fund forward.

\begin{figure}[t]
\centering
\includegraphics[width=1.0\linewidth]{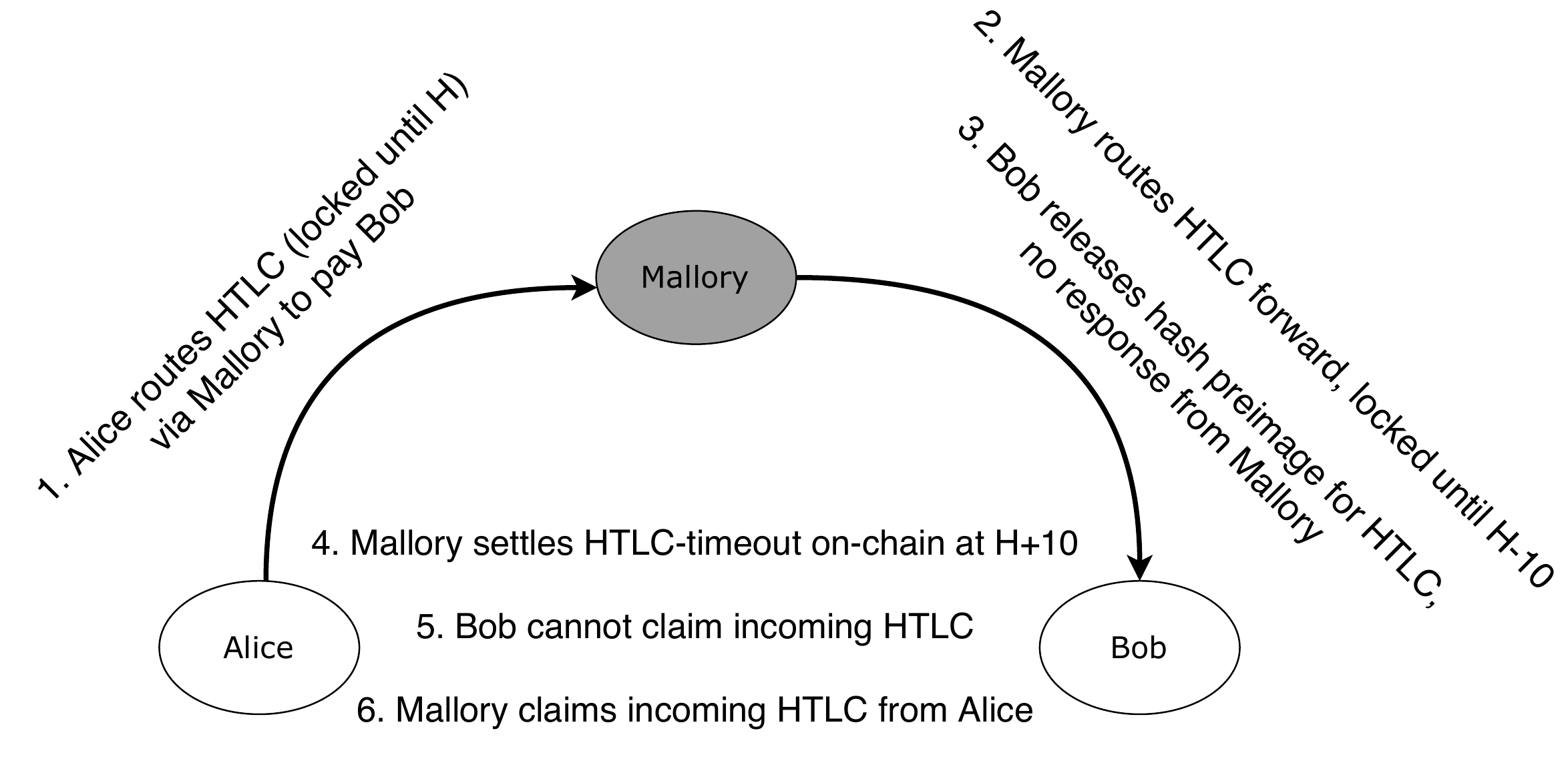}
\caption{Attack A3. Mallory is an attacker, Bob is a time-dilated victim, Alice simply routes a payment via Bob.}
\label{fig:a3}
\end{figure}

This attack differs from the previous one because an attacker only needs one channel with a victim, but also needs to be selected for a payment path. It also differs in the way an attacker finalizes the channel on-chain while stealing funds (by timing out a stolen HTLC). We summarize the attack in Fig. \ref{fig:a3}.

\section{Evaluation}
\label{sec:evaluation}

The practicality of time-dilation attacks once the node is already eclipsed can be measured by the \textit{time it takes to perform them} and the \textit{failure rate}. The timing aspect is relevant because if the attack takes several days, it can be more easily disrupted by a random event (e.g., a scheduled restart making a victim node connect to new peers).

Both of these metrics depend on the countermeasures software uses to disrupt time-dilation attacks. We previously mentioned that there is currently no \textit{dedicated} countermeasure implemented for this purpose. In this section we explore how one mechanism, originally designed for another purpose, bounds the practicality of time-dilation attacks.

The only heuristic employed by Bitcoin Core implementation which may help to break free from eclipsing is \textit{stale tip detection}. If a block hasn’t arrived during the last 30 minutes, a node attempts to establish one extra outbound connection and sync tips with a new peer, and then repeats it in 10 minutes if a block was not found during that time. This feature was originally introduced to handle \textit{non-malicious} failures of honest nodes to provide the latest blocks.

This countermeasure does not guarantee to mitigate the Eclipse attack, because a victim’s chosen extra outbound connections may be the attacker’s Sybil node.

We will further refer to the probability of this event as a \textit{failure rate}, upon which an attacker can improve. For example, an attacker can degrade the effectiveness of this extra connection by poisoning the victim's address manager while the node is eclipsed. Although it was demonstrated to be impractical \cite{Heilman2015Eclipse, Tran2019Erebus}, it may become feasible when a node is eclipsed.

\subsection{Optimal Attack Strategy}
To reduce the possibility of the victim’s Bitcoin node de-eclipsing due to the stale tip detection, the attacks we demonstrate  never intentionally trigger stale tip detection. An attacker should never exceed 30-minute delay between delivering blocks. It is possible, however, that stale tip detection may be triggered naturally by the randomness of the block mining process. According to our estimates, this happens with a probability of 5\% ($e^{-(30/10)}$), so on average 7 times a day.

The optimal strategy for an attacker would be to delay every block by 29.5 minutes. This time value is chosen to never exceed a 30-minute stale tip detection threshold with a room for network and processing latency.

This approach works best because it allows the fastest time-dilation without triggering the stale tip check. At the same time, it reduces the probability of "natural" de-eclipsing as much as possible: an attacker accumulates the time which can be used to amortize naturally slow (>30 minutes) blocks in the most efficient way. If an attacker combines it with address manager poisoning to make sure de-eclipsing doesn't help the victim, the delay can be increased.

We created a simulation-based model accounting for the exponential nature of block generation and stale tip detection. In our model, we simulate a generation of 1,000 blocks, and model an attack per which attacker delays \textit{every} block by a constant chosen interval of 29.5 minutes, so that the stale tip check is never triggered. We repeat this experiment 100,000 times for every configuration.

\subsection{Length and Success Rate of the Attacks}
According to our model, the time it takes to become ahead of a victim by 144 blocks is 36 hours.
We summarize the estimated time of keeping a node eclipsed required to perform time-dilation attacks based on the configurations of Lightning Network implementations in Tables \ref{tab:timings_a1}, \ref{tab:timings_a2}, \ref{tab:timings_a3}. These results are based on the configurations presented in Table \ref{tab:config_impl}.

\begin{table}[]
\begin{tabular}{l|l|l|l|l}
 & Implementation & Eclipse time (N) & Eclipse time (BC) &  \\
 \hline
 & C-lightning    & 24h     & 36h            &  \\
 & LND            & 24-336h & 36-508h &  \\
 & Eclair         & 120h    & 182h   &  \\
 & Rust-lightning & 24h     & 36h  &
\end{tabular}
\caption{\label{tab:timings_a1} The time a node has to remain eclipsed to allow attack A1 with Bitcoin Core and Neutrino backends.}
\end{table}

\begin{table}[]
\begin{tabular}{l|l|l|l|l}
 & Implementation & Eclipse time (N) & Eclipse time (BC) &  \\
 \hline
 & C-lightning    & 2.3h & 4h  &  \\
 & LND            & 6.6h & 10h &  \\
 & Eclair         & 24h  & 36h &  \\
 & Rust-lightning & 12h  & 18h &
 \end{tabular}
\caption{\label{tab:timings_a2} The time a node has to remain eclipsed to allow attack A2 with Bitcoin Core and Neutrino backends.}
\end{table}

\begin{table}[]
\begin{tabular}{l|l|l|l|l}
 & Implementation & Eclipse time (N) & Eclipse time (BC) &  \\
 \hline
 & C-lightning    & 1.2h     & 1.9h            &  \\
 & LND            & 1.7h & 2.7 &  \\
 & Eclair         & 1.9h    & 2.9h   &  \\
 & Rust-lightning & 1h     & 1.6h  &
\end{tabular}
\caption{\label{tab:timings_a3} The time a node has to remain eclipsed to allow attack A3 with Bitcoin Core and Neutrino backends.}
\end{table}

To confirm the results produced by the model, we derived an intuitive formula. The following formula assumes that every block comes exactly in 10 minutes, instead of the exponential distribution observed in practice. This is why the formula does not include any failures (a block interval can't exceed 30 minutes). The formula computes how long an attacker has to keep a victim eclipsed ($ET$), if an attacker needs to be ahead of the victim for a given number of blocks ($TL$) to break the timelock, and an attacker is limited to delay every block only for a given time ($SR$).

Given a targeted timelock($TL$, in blocks) and a per-block malicious slowdown rate ($SR$, in minutes), an attacker can estimate the required eclipsing time ($ET$, in minutes):
\begin{equation}
ET=(TL + \frac{10}{SR} * TL) * 10
\end{equation}

The formula intuitively reads as follows: $TL$ is the number of blocks of advantage required to be mined in the network, $\frac{10}{SR}$ represents how much a victim's blockchain tip moves per every \textit{block of advantage}, while an attacker dilates them by $TL$ blocks. The blocks in the second term are "undilatable" and have to be mined at a normal rate because all the dilation went into producing blocks in the first term. Since both of these values are in blocks, we need to multiply by 10 to get a result in minutes.

For Neutrino, where $SR$ is unbounded, an attacker has to simply wait when an honest network mines $TL$ blocks, because the second term is zero, the victim's tip just never moves. For $SR$=0, the result is infinity, meaning it's impossible to perform the attack without any slow down rate.

Let's say an attacker wants to be ahead of a victim by $TL$=40 blocks, and they can dilate at $SR$=0.33h per block (20 minutes). In this case, they would have to eclipse a victim for $ET$=3.33h. This example corresponds to the case of exploiting CLTV timelock of LND, which the model claims to be possible within 10 hours with a slowdown rate of 30 minutes, as set by the Bitcoin Core constraints we previously discussed.

Stale tip check can be triggered naturally even under the most optimal attack strategy. It would happen when it took \textit{very long} to mine a block. In this case, the \textit{delayed delivery time} of a particular block will be behind the actual block generation time. According to our model, with a chosen strategy of delaying for 29.5m every time, the probability of this natural de-eclipsing (attack failure rate) is around 7\%.

Intuitively, the probability of successful de-eclipsing via the stale tip check rapidly goes down with every maliciously delayed block. For a first block to trigger a stale tip detection (while a node is under attack), the natural mining time of that block should exceed 30 minutes, while for a fourth block it should exceed 80 minutes.
The probability of these events is 5\% and 0.03\% respectively.

Since Neutrino does not implement stale tip detection, there is no such upper bound, and the time it takes to dilate a node by a chosen number of blocks is constrained only by the natural time to produce those blocks. At the same time, without this check, the attacks on Neutrino \textit{always} succeed.

If an attacker had (or chose) to use 19.5m delays instead of 29.5m, it would increase the attack failure rate from 7\% to 22\%, while increasing the time it takes to perform time-dilation from 25-32h for reaching a difference of 100 blocks. Reducing the stale tip threshold wouldn't help against time-dilation, because it would significantly increase false positives.

\subsection{Gain from the Attacks}

Even though Eclipse attacks against full nodes are difficult expensive \cite{Tran2019Erebus}, an attacker may steal all liquidity available at a victim's LN node at once. In addition, since latest Eclipse attacks are infrastructure-level \cite{Tran2019Erebus, Apostolaki2017Hijack}, they may target several nodes at once. Since full nodes are often used by large LN hubs or service providers and have high available liquidity, an attacker may justify the high Eclipse attack cost by stealing aggregate liquidity from several nodes at once. As for the LN users running light clients, stealing from them is already easy enough, so even stealing rather low amount may be justified.

The amount per channel to be stolen from a victim is technically different across attacks, but usually equals the channel capacity. Lightning Network recently lifted \cite{Bolt2020Wumbo} the channel capacity bounds, allowing users to open as large channels as they want. In practice, a median LN node capacity is 0.003 BTC, and the total amount locked in LN channels is 940.5 BTC, as of May 2020\footnote{1ml.com}.

\textbf{Attack A1} assumes that an attacker commits one of the outdated states on-chain. To maximize the gain, an attacker would claim the state where they had the largest amount. Picking a particular state does not make an attack more difficult. An attacker can then steal \textit{full channel capacity}, minus a small value ("channel reserve"), enforced by the protocol to disincentivize channel revocations.

It is possible that there was no state per which all funds were located on the attacker's side of the channel. In that case, an attacker can route payments to themselves via the victim, and move funds to the attacker's side of the channel.

\textbf{Attacks A2 and A3} rely on stealing in-flight HTLC, so they can steal at most the maximum in-flight value, as negotiated during channel opening. In most of the LN implementations, this value is by default the same as the full channel capacity.

\section{Countermeasures}
\label{sec:countermeasures}

We split countermeasures into preventing time-dilation itself, preventing the exploitation of it, and the issues related to reacting once the exploitation is detected. The ideas we suggest are heuristics: they can't guarantee full mitigation of time-dilation attacks, but rather increase the attack cost.

At the end of this section we discuss WatchTowers separately, because they can be used as all three types of countermeasures.

\subsection{Preventing Time-Dilation}
\label{sec:preventing_td}

\textit{Preventing Eclipse attacks} on Bitcoin nodes would make time-dilation impossible. The cost of Eclipse attacks can be increased via the following measures.

\textbf{Higher connectivity and the number of honest reachable nodes.} As it was shown in Section \ref{sec:eclipse_neutrino}, the probability of Eclipse attacks goes down when any of the two parameters are increased. There are three ways to achieve this:
\begin{itemize}
\item encourage users to provide more resources (bandwidth, computational) to the network
\item make the use of those resources more efficient
\item increase adoption and deployment of BIP 157 (especially server-side) across the ecosystem
\end{itemize}

\textbf{Peer diversity.} Since both Eclipse attacks and transaction origin inference involve an attacker connecting to the victim, increasing the cost of Sybil attacks is an effective countermeasure. If honest nodes make decisions based on some scarce property, it would make it more difficult for an attacker to gain enough connections. Complementing anti-Sybil mechanisms (e.g., peer diversification based on the peer’s Autonomous System) with proactive topology improvements through peer rotation would help to break free from ongoing Eclipse attacks \cite{Tran2019Erebus}.

\textbf{Link layer diversity.} A natural countermeasure to peer-to-peer layer attacks is communication redundancy: using several interconnected multi-homed nodes, a VPN, a mesh network, or the Lightning Network itself for block and transaction relay. If any of these methods are employed to receive blocks and transmit transactions, an attacker would be required to disrupt the attacks as well. in the case of bandwidth-constrained communication channels, transmitting block headers would be enough to detect an anomaly.

In addition, LN clients may serve each other to increase their security in a web-of-trust-style deployment. A "friendly" client may be asked to watch a list of outputs spending belonging to another client and notify in the case of a match against their filters. Therefore, an attacker would have to control all chain providers of every relevant swarm. Given resources and incentives for the watching client and the privacy leak for the beneficiary client, this scheme relies on social trust assumption (e.g., a set of mobile wallets belonging to a family).

\textbf{Peer-to-peer protocols anonymity} is a standalone research topic. Integrating ideas from prior work \cite{Fanti2018Dandelion++, Naumenko2019Erlay} into Bitcoin Core, as well as improving the existing features may make time-dilation attacks impractical.

Although \textbf{straightforward use of Bitcoin over Tor} was demonstrated to be vulnerable to certain attacks \cite{Biryukov2015Tor}, other designs of transaction relay mechanisms involving various mixnets should be explored.

\subsection{Detecting Time-dilation}

Although in Section \ref{sec:evaluation} we demonstrated that the current stale tip detection technique is limited, a specialized time-dilation detection could be useful.

The local system clock can be used to detect the absence of new blocks at an unlike-enough interval, similarly to the stale tip check, although local clock can be a subject to manipulation or system errors. Alternatively, time present in the header of a mined block may be compared to the local time, although this field in block headers is only moderately enforced by consensus. These methods can be expanded to consider a series of blocks.

\textbf{Lightning implementation-level warnings} in the case of the observed anomalies may help a node operator identify that a node is currently under attack. For example, if a Lightning node receives channel announcements related to blocks “from too far in the future”, it may issue a warning.

Similarly, \textbf{abnormal routing failure rate} may be used. If a Lightning node is behind in its Bitcoin blockchain view, but Lightning payments between honest nodes are still flowing through it, this node will have a high routing failure rate. This would happen because honest nodes on the routing path would reject the forwarded HTLC for being too close to expired. This observation can be used to detect time-dilation.

The implementation of these measures is not trivial. All these solutions have a fundamental trade-off: security against false positive detection rate. And even if an attack was properly detected, it is unclear what the victim's reaction should be. In Section \ref{sec:td}), we explored why this is problematic by looking at the currently implemented stale tip detection.

\subsection{Reaction}

Even if an LN node detected it is under a time-dilation attack and it's not too late, it still cannot easily prevent the loss of funds. The issues with stopping the attack include:
\begin{enumerate}
\item If there are multiple channels opened, it is unclear which of them should be closed to prevent the loss
\item For a given channel, it may be unclear which state was committed by the attacker. This must be taken into account by the victim when constructing a justice transaction.
\item A justice transaction should have a proper fee and be transmitted to the miners via the honest peer-to-peer network, or it won't be confirmed.
\end{enumerate}

The challenges (1), (2), (4) are critical in the context of a victim’s Bitcoin node being eclipsed. Thus, even if the attack was detected, the only solution is to apply the same anti-eclipsing mechanisms we suggested above.

\subsection{WatchTowers}

This approach implies that chain monitoring is replicated with different computers, each of them maintaining the channel state \cite{Dryja2016Watchtowers, khabbazian2019outpost, avarikioti2020cerberus, Mccorry2018Pisa}. WatchTowers provide an alternative stack operating over a separate infrastructure. This raises the bar for an attacker by making Eclipse attacks and transaction origin inference, and thus imrpoves on all three suggested countermeasure directions.

The current discussion around WatcTtowers usually assumes they are operated by special providers, and not users themselves. This increases the robustness even further but introduces an extra assumption about the WatchTower provider.

\section{Discussion}
\label{sec:discussion}

\textbf{Other time-sensitive protocols.}
While the attacks we demonstrated were specifically targeting Lightning Network, we believe that a wide variety of Bitcoin second layer protocols \cite{dryja2017discreet, LeGuilly2020OffDLC, moser2016bitcoin, Gibson2017Coinswaps} may be susceptible to time-dilation attacks. This applies to any of them where timelocks are used to arbitrate parties willing to commit concurrent on-chain transactions. We believe that designers of those protocols should take time-dilation threats into account whilst arguing about their security.

\textbf{Combined attack with mempool spam.}
The attacks we discussed may be prevented by the victim detecting it and submitting a justice transaction in a timely manner. During an attack, a victim has to act in a very limited time-frame, less than the one anticipated in the original timelock. An attacker may make it even harder for a victim by running a DoS against Bitcoin, to ensure the victim’s transaction is not confirmed. If LN implementations employ dynamic fee-bumping, this may help a victim to prioritize their transactions and overcome DoS.

\textbf{Attacker controls broader infrastructure.}
The attacks we demonstrated work under limited capabilities of the adversary. If an adversary controls the victim's ISP, or has ways to influence DNS responses, or have other ways to exploit the key infrastructure, the attacks may be executed at a much lower cost. It also makes countermeasures we suggested much less efficient.

\textbf{Better mapping techniques}
There are more advanced techniques for correlating Bitcoin and Lightning nodes. For example, an attacker can use timing analysis of bootstrap/restart or force a Lightning node to close a channel to speed up transaction origin inference. We leave this research for future work.

\textbf{Initial Block Download after 24h.}
Although we mentioned that there is currently no dedicated mechanism for time-dilation detection, one relevant feature of Bitcoin Core software is switching to Initial Block Download mode. This switching can then be detected by LN operator. It happens if the time defined in the latest known block header is 24h behind the current system time. This feature is not efficient against time-dilation attacks, because, as we demonstrate in this paper, they need to dilate less than 24 hours. We do not recommend modifying it to be useful in this context, because it was not originally designed to prevent attacks.

\textbf{Explore tradeoff between higher-security and fund liquidity.}
Increasing the timelocks would require an attacker to keep a victim eclipsed for longer, and would give a victim more time to prevent the attack or react to it. This makes more secure channels less dynamic because in a non-cooperative case it takes longer to settle them. Attack cost may be spread by exploiting all links of a single LN node, so the sum of all channel values should be used to assess operational risks. Reasoning on time-value only is an incomplete method to argue about sophisticated attacks.

Every LN operator should separately consider their own risks related to time-dilation, in addition to the time-value trade-off of payment channels. Finding the proper balance between the systemic risk caused by the \textit{liquidity market for routing payments} \cite{Bitmex2019RoutingFees} and security is an open area of research.

\section{Related work}
\label{sec:related_work}

Attacks on the Bitcoin peer-to-peer network usually result in eclipsing honest nodes and transaction deanonymization.

The first Eclipse attack on Bitcoin was based purely on the high-level Bitcoin network protocols \cite{Heilman2015Eclipse}, while the latter exploited BGP and Internet infrastructure \cite{Apostolaki2017Hijack, Tran2019Erebus}. The studied consequences of Eclipse attacks include monetary (double-spending attacks, attacks on mining) and non-monetary (peer-to-peer layer deanonymization). The prior art demonstrated that it's possible to steal funds via an Eclipse attack via on-chain double-spend, although it would require access to hashrate and purchasing something from a victim. Thus, it is more difficult than with time-dilation attacks.

The consequences of Eclipse attacks for the second layer were only briefly mentioned \cite{Heilman2015Eclipse}. Bitcoin was also considered as a target for attacks on NTP \cite{Malhotra2015NTP, Herrera2019HidingBalance}, although the consequences for the second-layer protocols were also not explored.

Attacks on the privacy of the Bitcoin peer-to-peer protocols demonstrated that transaction deanonymization is fairly feasible both with simple techniques like first-spy estimator and more advanced strategies \cite{Fanti2017Dandelion, Fanti2018Dandelion++, Biryukov2014Deanonymisation, Naumenko2019Erlay, biryukov2019deanonymization}.

Some of the privacy improvements were deployed, but they can't address the issues in full. Using Tor at the peer-to-peer layer was demonstrated to be an inadequate solution to these problems \cite{Biryukov2015Tor}.

Attacks on the Lightning Network may be split into two groups: privacy-related or DoS-related. Prior work on privacy mainly explored real-time balances of channels in the network via probing \cite{Tikhomirov2020ProbingBalance}. These DoS attacks achieve cheap network congestion preventing the flow of honest payments \cite{perez2019lockdown, rohrer2019discharged, mizrahi2020congestion}. In light of our work, an attacker may use route hijacking techniques \cite{tochner2019hijacking} to prepare targeted channels for time-dilation.

It was also explored how an attacker can steal routing fees \cite{malavolta2019anonymous}, and exploit transaction propagation policies to get an advantage in LN settlement \cite{Corallo2020Pinning}.

To the best of our knowledge, none of the major LN update proposals \cite{Decker2015Duplex, Decker2018Eltoo, Poelstra2019PTLC} can solve the time-dilation issues.

\section{Conclusions}
\label{sec:conclusion}

Even though Lightning Network has the potential to address the scalability limitations of Bitcoin, it introduces new security assumptions. In this work we explored how they hold in practice.

More specifically, we explored what can be done when an attacker isolates (eclipses) a user of the Lightning Network, and feeds blocks to the victim at a slower rate. We showed that time-dilation cannot be addressed by simply detecting slow block arrival, and implementing sophisticated detection measures is not trivial.

We argued that time-dilation attacks are currently the most practical way of stealing funds via Eclipse attacks since time-dilation attacks do not require access to hashrate and an attacker doesn't have to purchase anything from a victim. The Eclipse attack cost can be justified against both light clients (the cost is low) and full nodes (an attacker may steal all liquidity of wealthy nodes at once).

Finally, we suggested that strong anti-Eclipse/anti-Sybil measures (e.g. alternative sources of blocks) is the key to significantly reducing the risks of time-dilation attacks.
\section*{Acknowledgments}
\label{sec:acks}

We are grateful to Suhas Daftuar and Matt Corallo for initial discussions on time-dilation attacks.
We would like to thank Clara Shikhelman, Sergei Tikhomirov, Rene Pickhardt, Sergi Delgado Segura, Michael Folkson, James Chiang, devrandom, Bastien Teinturier and Thibaut Le Guilly for the useful feedback on the paper.

This work was supported by Chaincode Labs.
\bibliographystyle{apa}
\bibliography{paper}
\end{document}